\documentclass[preprint,showpacs,preprintnumbers,amsmath,amssymb]{revtex4}

\usepackage[dvips]{graphicx}
\usepackage{amsfonts}
\usepackage{amsmath}
\usepackage{longtable}

\begin{document}

\title{Calculation of accurate permanent dipole moments of the lowest $^{1,3} \Sigma^+$  states 
of heteronuclear alkali dimers using extended basis sets}

\author{M. Aymar}
\author{O. Dulieu}
 \email{olivier.dulieu@lac.u-psud.fr}
\affiliation{Laboratoire Aim\'e Cotton, CNRS,
 B\^at. 505, Campus d'Orsay, 91405 Orsay Cedex, France}

\date{\today}

\begin{abstract}
Obtaining ultracold samples of dipolar molecules is a current challenge which requires an 
accurate knowledge of their electronic properties to guide the ongoing experiments. In this paper, 
we systematically investigate the ground state and the lowest triplet state of mixed alkali dimers 
(involving Li, Na, K, Rb, Cs) using a standard quantum chemistry approach based on pseudopotentials 
for atomic core representation, gaussian basis sets, and effective terms for core polarization 
effects. We emphasize on the convergence of the results for permanent dipole moments regarding the 
size of the gaussian basis set, and we discuss their predicted accuracy by comparing to other 
theoretical calculations or available experimental values. We also revisit the difficulty to compare 
computed potential curves among published papers, due to the differences in the modelization of 
core-core interaction.
\end{abstract}


\maketitle

\section{Introduction}

Researches on ultracold molecules are progressing very fast since a couple of years, as a
 growing number of research groups are now involved in this field. Indeed, ultracold molecules
  offer many new opportunities compared to ultracold atoms, due to their more complex structure: 
they can carry an amount of internal energy larger by orders of magnitude than their 
 kinetic energy, opening new areas like ultracold photochemistry \cite{dulieu2004} or also superchemistry 
 \cite{heinzen2000,olsen2004}. The ultimate control of elementary chemical reactions using short-pulse lasers is now 
 foreseen \cite{koch2004a}, as both the internal and external degrees of freedom of the molecules can be 
 well mastered. They allow the reinvestigation of phase transitions which up to now were 
 explored only from the point of view of condensed matter physics, like the transition from 
 bosonic to fermionic statistics \cite{zwierlein2004,regal2004,bartenstein2004}, the superconductivity transition
through the observation of the pairing gap in strongly interacting Fermi gases \cite{chin2004a}
 or the superfluid transition via the study of the BEC-BCS crossover \cite{bartenstein2004a}. Spectacular 
 achievements have been recently demonstrated as for instance the observation of molecular 
condensates with various alkali dimers \cite{herbig2003,greiner2003,zwierlein2003}, or of Feshbach resonance
in molecule-molecule ultracold collisions \cite{chin2005}.
 
Other interesting prospects concern dipolar molecules (ie with a permanent dipole moment), as it can 
be read through the special issue on this topic published in 2004 \cite{epjdspecial}. Among those are 
the possibility to design a quantum information device \cite{demille2002}, or the test of fundamental 
theories like the measurement of the eletron dipole moment \cite{kozlov2002,hudson2002}.
 
The formation process is obviously the key issue of these developments. An efficient approach 
which dramatically improved these last years is the so-called Stark deceleration technique, 
relying on the slowing and trapping of molecules with a permanent electric dipole moment 
(usually labelled as dipolar molecules) using inhomogeneous external electric fields \cite{bethlem1999}. 
Besides, the "`historical"' approach based on photoassociation of ultracold atoms which has
 first demonstrated the formation of ultracold molecules with cesium dimers \cite{fioretti1998}, has 
 recently progressed with the creation of ultracold samples of dipolar molecules composed of 
different alkali atoms like RbCs \cite{kerman2004,sage2005}, KRb \cite{mancini2004,wang2004}, and NaCs 
\cite{shaffer1999a,haimberger2004}. 
 
In a previous paper \cite{azizi2004} we computed the rates for the photoassociation of mixed alkali pairs, 
and for the susbsequent formation of cold molecules, which showed that all alkali pairs 
involving either Rb and Cs are well suited  for that purpose, as the cold molecule formation 
rate was only about ten times smaller than for Cs$_2$ formation. This result was found in
 good agreement with the RbCs experiment of Kerman {\it et al} \cite{kerman2004}. Potential curves 
available in the literature, and constant (atomic) transition dipole moment were proved to be sufficient 
to establish these estimates. However, the practical implementation of cold molecule formation
 via photoassociation will require a much better knowledge of their electronic properties like 
radial variation of permanent and transition dipole moments,
in order to better guide the experimentalists towards specific systems and transitions with 
maximal efficiency. New spectroscopic studies on heteronuclear alkali pairs are also currently 
reinvestigated for NaRb \cite{knoeckel2005} and NaCs \cite{docenko2004}, as well as the investigation of the effect
of strong electric fields \cite{gonzalez-ferez2004}, which may also benefit of such a study. 

To respond to these requirements, we started a new accurate analysis of electronic properties 
of all alkali pairs from Li to Cs, including potential curves for ground and excited states, 
permanent and transition dipole moments. 
We set up an automatic procedure based on the CIPSI package (Configuration Interaction 
by Perturbation of a multiconfiguration wave function Selected Iteratively) \cite{huron1973}
developed by the "Laboratoire de  Physique Quantique de Toulouse (France)". Here we first investigate 
the permanent dipole moment of the ground state and the lowest triplet state of all mixed alkali
 pairs, for which only scattered experimental or theoretical results are available. We display their
variation with the interatomic distance, as well as with the vibrational level. Most of these results 
were not previously available elsewhere. We emphasize 
on their convergence with respect to the size of the basis set. In the next section, we briefly 
recall the calculation procedure, which is well documented in many publications. We also review 
extensively the available calculations on these systems. In section \ref{sec:results}, we present 
our results obtained with basis sets published in the literature, and with new extended basis sets 
designed for all alkali atoms, in order to address convergence issues with respect to the size of the
basis. We also discuss the comparison with other published papers, which 
is not often proposed, especially in the perspective of the addition of effective terms to the computed 
energies.  

\section{Theoretical calculations}
\label{calcul}

Many studies on heteronuclear alkali dimers have been performed using various quantum chemistry methods, 
restricted to the calculation of electronic potential curves for the ground and excited molecular states. 
The CIPSI package
has been used to compute ground and excited potential curves of RbCs 
\cite{pavolini1989,allouche2000a,fahs2002}, NaK \cite{magnier1996,magnier2000}, 
KLi \cite{rousseau1999}, LiRb, NaRb \cite{korek2000a}, LiCs, NaCs, KCs \cite{korek2000}, 
KRb \cite{rousseau2000a}. The method is based on the {\it ab initio} pseudopotentials
of  Durand and Barthelat \cite{durand1974,durand1975} for core representation,
phenomenologic $\ell$-dependent effective core polarization potential \cite{foucrault1992}
and self-consistent calculation (SCF) combined with full valence configuration interaction 
interaction (CI) calculations. Additional effective terms may also be used to take in account 
short range core polarization terms (see for instance ref. \cite{pavolini1989}).
Potential energy curves of ground  and excited states of LiNa have been investigated  
by Davies {\it et al} \cite{davies1981} through many-body perturbation theory calculations, and by
Schmidt-Mink {\it et al} \cite{schmidt-mink1984} who have combined all-electron  (SCF) calculations, 
valence CI calculations and have included an effective core polarization potential
\cite{muller1984}. The electronic structure of KRb has been studied by 
Leininger {\it et al} \cite{leininger1995} and by Park {\it et al} \cite{park2000}
 using a method based on restricted Hartree-Fock (RHF) calculations, using the small-core 
(with nine electrons for each atom) relativistic pseudopotential of Christansen {\it et al}
 \cite{hurley1986} and CI calculations. Park {\it et al} also introduced the effective
core polarization potential of ref.\cite{muller1984a}. Note that an extensive bibliographic
overview on theoretical as well as experimental papers can be found in  the "DiRef" database
 \cite{bernath2001} devoted to diatomic molecules. 

Only few papers include theoretical values for permanent dipole moments in addition to 
potential curve calculations. Igel-Mann {\it et al} \cite{igel-mann1986} have performed a
systematic study of the 
permanent dipole moments around the equilibrium distance of the ground state of all alkali 
dimers using the semi-empirical pseudopotential of Fuentealba \cite{Fuentealba1982} including 
a core polarization potential.  All-electron SCF and valence CI  calculations with or without 
core polarization correction have been performed by M\"uller and Meyer \cite{muller1984}
for LiNa, LiK and NaK at the equilibrium distance.  Janoschek {\it et al} \cite{janoschek1978}
 and Stevens {\it et al} \cite{stevens1984} have reported permanent dipoles of NaK.
 The {\it ab initio} and pseudopotential  calculations of Janoschek \cite {janoschek1978} concern 
only the ground state, while the  more complete  SCF and CI calculations including core-polarization 
effects done by Stevens {\it et al} \cite{stevens1984}  include
the $R$-dependence of the ground and triplet state permanent dipole moment.
SCF and CI calculations of  permanent dipole moments for NaLi  have been
performed by Bertoncini {\it et al} and by Rosmus and Meyer 
\cite{bertoncini1970,rosmus1976}. Bertoncini {\it et al} reported
 the  $R$-dependence of the dipole in the ground  and  the lowest triplet states. 
 
{\it Ab initio} calculations of the $R$-dependent dipole moments
 of the $^{1,3} \Sigma^+$  states of KRb have  been performed by Kotochigova {\it et al}
\cite{kotochigova2003}, using  Hartree-Fock or Dirac-Fock orbital basis set complemented by
Sturmian functions to describe virtual orbitals and CI valence-bond calculations.
Core-valence correlations are accounted by allowing single electron excitation from the closed
$3p^6$ of K and $4p^6$ of Rb  and by introducing several virtual orbitals. Core polarization 
effects are not explicitely mentionned by the authors in their papers. They found a large
influence of relativistic effects on the dipole  moments, with an increase of about 20\% for
the ground state and of about 50\% for the a$^3\Sigma^+$ case, in the region of the potential
minimum.

Besides, only a few experimental values for the permanent dipole moments of the ground
state obtained by different techniques have been reported.
Dagdigian and Wharton \cite{dagdigian1971,graff1972,dagdigian1972} used molecular beam electric
 deflection and resonance spectroscopy to determine the molecular electric dipole of several
heteronuclear alkali dimers. The permanent dipole of NaLi or KLi molecules have been determined
by Engelke {\it et al} using laser induced fluorescence \cite{engelke1982,engelke1984}, and
the one of  NaK by Wormsbecher {\it et al} using microwave optical double resonance technique
 \cite{wormsbecher1981}. Finally, Tarnovsky {\it et al} \cite{tarnovsky1993} have measured 
the electric dipole polarizabilities of alkali homonuclear and  NaK and KCs heteronuclear
dimers using molecular beam deflection in an inhomogenous electric field. This 
 led the authors to
set up two empirical rules for determining polarizabilities and dipole moments of all
heteronuclear alkali dimers, yielding then the values of the dipole moments for all mixed pairs
from the difference between the polarizabilities of their respective homonuclear dimers.

As mentionned above, we use here the CIPSI package to investigate the electronic structure of mixed 
alkali pairs MM', and we recall now the main features of this code. In this method, the atomic cores are 
described by the pseudopotentials of  Durand and Barthelat \cite {durand1974,durand1975}
whose  parameters have been adjusted to reproduce the energies and valence orbitals of all-electron 
Hartree-Fock  SCF calculation for the atomic ground state. For Rb and Cs, the pseudopotential includes 
the mass-velocity, and the Darwin relativistic corrections. Each atom is described by a Gaussian basis set.
As the alkali atoms are treated as one-electron species, correlations are not explicitly introduced.
Core polarization is taken in account through an $\ell$-dependent effective core 
polarization potential (ECP) \cite {foucrault1992} depending on the dipole polarizability
$\alpha_d^{M^+}$ and $\alpha_d^{M'^+}$ of the M$^+$ and M'$^+$ ions and of $\ell$-dependent cut-off parameters 
$\rho_{\ell}$ adjusted to reproduce the experimental energies \cite{moore1949,moore1952,moore1958}
 of the two lowest $s$, $p$ and $d$ atomic levels. Molecular orbitals are then determined by restricted 
Hartree-Fock calculations and full valence CI are performed.

Additionnally, the potential energy should include the core-core interaction potential $V_{cc}(R)$ which is 
modelled as long as the cores are far enough from each other with the pure repulsive $1/R$ term,
and an attractive charge-induced dipole term $V_{cc}^{ind}(R)=-(\alpha_d^{M^+}+\alpha_d^{M'^+})/2R^4$. 
When the cores come close to each other, such an 
approximation is not sufficient due to their strong electrostatic repulsion, and further short-range effective 
terms have to be included in $V_{cc}$. As discussed for instance in the case of alkali dimers  
\cite {pavolini1989,jeung1997,rousseau1999a}, the core-core repulsion $V_{cc}^{rep}(R)$ has to be evaluated from ab-initio 
calculations according to various assumptions. The frozen core approach is then often used, in which the repulsion 
between the two ionic cores is obtained from a SCF calculation where the two ionic cores are represented by frozen 
atomic orbitals. It has been shown \cite{pavolini1989} that this term can be fitted by an exponential form. The 
contribution of the core-core dispersion energy has also to be taken in account, approximated by the London formula 
(see for instance ref.\cite{pavolini1989}):

\begin{equation}
V_{cc}^{disp}(R)=-\frac{3\alpha_d^{M^+} \alpha_d^{M'^+}}{2R^6} \frac{E_I^{M^+} E_I^{M'^+}}{E_I^{M^+}+E_I^{M'^+}}
\label{eq:disp}
\end{equation}

where $E_I^{M^+}$ and $E_I^{M'^+}$ are the ionisation energies of the M$^+$ and M'$^+$ ions respectively.

We performed three different series of calculations, described below, characterized by different 
Gaussian basis sets summarized in Tables \ref{tab:base}, \ref{tab:exponents-cont}, and \ref{tab:cutoff}. The resulting
atomic energies are displayed in Table \ref{tab:atomic}, while the size of the generated molecular basis set is 
given in Table \ref{tab:base2} for mixed alkali pairs.

\begin{itemize}
\item Prior to our new calculations, we checked the code and our procedure by first reproducing the computations
performed by previous authors, employing the contracted Gaussian basis sets, the ionic core dipole polarizabilities 
and the $\ell$-dependent cut-off radii $\rho_{\ell}$ published by Poteau {\it et al} \cite{poteau1995} for Li,
by Magnier {\it et al} \cite{magnier1993} for Na, by Magnier {\it et al} \cite{magnier1996} for K, and by Pavolini
{\it et al} for Rb and Cs \cite{pavolini1989}. Note that in these calculations, the values of 
$\alpha_d$  may come from different sources \cite{wilson1970,muller1984}. These calculations will be refered to as
 belonging to the "A" series. 

\item Then we removed the contracted orbitals from the basis
sets quoted just above, and introduced the $\ell$-dependent ECP's afterwards. For all atoms, we obtained a better
description for atomic energy levels than with the contracted basis sets. Guided by this trend, our next series 
of calculations (labelled as series "B") involve uncontracted Gaussian basis extended with respect those used in
 the "A" series, and dipole polarizabilities 
all coming from the same paper by Wilson {\it et al} \cite{wilson1970}. Some 
basis coefficients, and the cut-off radii, have been reoptimized, in order to improve the quality of calculated 
atomic energy levels. Note that the Li basis set has been considerably 
increased compared to ref.\cite{poteau1995}. In contrast, the K basis set is the same than in ref.\cite {magnier1996},
but  without contraction coefficients. The K basis has been recently extended by Magnier {\it et al} \cite{magnier2004} to
study highly excited states of the K$_2$ dimer.

\item Due to the current interest for cold molecule studies about heteronuclear alkali pairs involving Cs,
\cite{kerman2004,haimberger2004}, we set up a third series of computations (C series) for LiCs, NaCs, KCs, 
and RbCs, using a basis set and cut-off parameters for Cs,
which will be illustrative for the discussion of convergence issues.
\end{itemize}

As seen in Table \ref{tab:atomic} we obtained in most cases slight improvements on the atomic energy levels when
A and B calculations are compared to the experimental ones, which confirms that previously published basis sets 
were already quite well optimized. However, as it will be discussed in the next section, the improvement is more 
spectacular for potential curves, due to the cumulative effect of better defined static polarizabilities, and extended 
basis sets.

\section{Results}
\label{sec:results}

Before discussing in detail our results on permanent dipole moments, we first check the reliability of our extended
basis sets by looking at the potential curves obtained for the ground state and lowest triplet state of the 
mixed pairs, concentrating on two specific examples, RbCs and KRb. We first first look at the potential
energies $V'(R)$ including only the $1/R$ term of $V_{cc}(R)$. As it is shown in Figure \ref{fig:rbcspot_Xa},
the minimum energy for $V'(R)$ curves is found in general at distance $R_m$ larger than the equilibrium distance 
$R_e$ of the full $V(R)=V'(R)+V_{cc}^{ind}$ curves. The combined influence of the addition of diffuse orbitals and the 
improved adjustement of the
cut-off radii in series B compared to series A is clear, as the $X$ and $a$ states are found 48~cm$^{-1}$ and 
6~cm$^{-1}$ deeper than in series A, respectively. In the latter case, the basis C for Cs still increases the $a$ well
depth by 6~cm$^{-1}$, while no effect is visible for the X state. The same trend is observed for all pairs: as a further 
example, the KRb ground state is found in the B case about 70~cm$^{-1}$ deeper than in the A case, while the depth 
of the $a$ state is increased by 2~cm$^{-1}$. Excited potential curves are also modified from A to B calculations, and
will be investigated in further work.

We then computed the permanent dipole moment of $X$ and $a$ states for all heteronuclear pairs.
The sign of the permanent dipole moment of a molecule depends on the interatomic axis orientation, which is assumed 
here to be oriented along the MM' direction, where M is the lightest atom of the mixed pair. A negative dipole 
moment then implies an excess electron charge on the M atom. In the following, values are given in Debye, with 
1 atomic unit=2.54158059 Debye.

Here again we first check the influence of the optimization of the basis set 
and of the cut-off parameters on the RbCs permanent dipole moment 
(figure \ref{fig:rbcsdip_Xa}). Note that $V_{cc}(R)$ does not influence the 
electronic wave functions, neither
the dipole moments functions. The $R$ dependence is found very similar for all A, B, C calculations, with a minimum
around 9.3$a_0$ for the $X$ state (slightly larger than the minimum distance $R_m^X=8.5a_0$), and around 13.4$a_0$
for the $a$ state (slightly larger than the minimum distance $R_m^a=12.6a_0$). Increasing the size of the basis 
changes the dipole moment of the $X$ state in the region of the minimum by only 2 to 5\%, while the moment for 
the $a$ state, which is much smaller in magnitude, changes (around the location of its minimum) by about 50\% from 
A to B calculations, and only by 7\% from B to C calculations. A similar trend is observed for the other mixed
pairs (see table \ref{tab:compar_cont}), as the change from A to B or C approaches in the ground state dipole moment 
never exceeds a few percents, being most often smaller than 1\%. So we can safely estimate that the B calculations
are converged with respect to the size of the basis set.

The $R$-variation of the permanent dipole moment for the $X^1\Sigma^+$ and $a^3\Sigma^+$ states deduced from 
calculations B (figure \ref{fig:dipole_R}a) is found similar for the ground state of all pairs.
Apart for the LiNa dipole moment which remainsclose to 0 Debye, the curves present a minimum at a distance $R_d^X$
systematically larger by about 1$a_0$ to 1.5$a_0$ than the equilibrium
distance $R_e^X$ of the ground state, and vanishes at large $R$ as expected.
The magnitude of the dipole strongly depends of the chosen pair, the largest being for LiCs around 
-6~Debye. In contrast, the $R$-dependence of the dipole moment in the $a$ state is more irregular over
the mixed pair series, with  a minimum for some species (LiCs, RbCs, KRb) and a maximum for the other dimers. 

Figure \ref{fig:dipole_v} shows the variation of the permanent dipole moment with the vibrational level of the $X$ 
and $a$ states, obtained after averaging the dipole moment function over the vibrational wave functions of the 
potential functions $V(R)=V'(R)+V_{cc}^{ind}$. This figure partly reflects the $R$-dependence drawn in Figure 
\ref{fig:dipole_R}: the ground state dipole moment is almost constant for the lowest forty levels for the less
polar species, while it varies more rapidly for LiRb, LiCs, NaRb. As expected, all these curves vanish at large
distances (not shown on the figure). 

For the purpose of comparison with other works, we summarize our results for the ground state in a tabular form
(Table \ref{tab:compar_cont}), displaying the value of the permanent dipole moment at the equilibrium distance $R_e^X$,
at its extremum value $R_m^X$, and averaged for the $v=0$ vibrational level. To the best of our knowledge, the other 
available theoretical values for the permanent dipole moment of the mixed alkali pairs concern the ground state 
at its equlibrium distance $R_e^X$. In contrast, only the value of the permanent dipole
moment for a given vibrational level of molecules is obtainable in an experiment, limited to $v=0$ up to now.
Note that for simplicity we reported absolute values in table \ref{tab:compar_cont}: indeed, the experimentalists cannot 
determine the sign of the dipole moment, while it is not always clear in the theoretical papers what is the choice of the 
authors for the orientation of the molecular axis. 

As expected, the dipole moment of 
$v=0$ is close to the value of the dipole function at the equilibrium distance. Our results are in good agreement 
with the available experimental values, within a 2\%  for LiK, NaK, and NaCs, and slightly more (6\%) for NaRb. The
difference is found much larger (around 20\%) for the less polar molecule LiNa. 
We confirm that the approximate scaling law proposed in ref.\cite{tarnovsky1993} overestimates the dipole moment, 
except for the less polar molecules KRb and RbCs, which in contrast are underestimated. 
Our results are in excellent agreement with those of ref.\cite{igel-mann1986}, which a similar approach without the $\ell$-dependence of the core polarization terms. 

We have already mentionned that very few other theoretical variations of the permanent dipole moment
 with the interatomic distance
are available in the literature. The $R$-dependence of the dipole moment in the ground state of NaLi
molecule obtained by Bertoncini {\it et al} \cite{bertoncini1970} and by Rosmus and Meyer 
\cite{rosmus1976} disagree with each other both in sign and in magnitude. Our negative value for LiNa supports 
the prediction  of Rosmus and  Meyer but not their $R$-variation. Note that the authors of ref.\cite{rosmus1976}, and 
later on in another paper \cite{muller1984a}, underlined the great sensivity of the dipole moment of NaLi to
valence correlations and core polarization effects. Similarly, results by Stevens {\it et al} on NaK \cite{stevens1984}
 also differ strongly from ours, which can be attributed to the absence of an $\ell$-dependence in the effective core
potentials, included afterwards in the NaK study of Magnier and Milli\'e \cite{magnier1996}. 

The KRb molecule is an attractive case in the cold molecule field, as two recent 
calculations of the $X$ and $a$ states of KRb including the $R$-dependence of the permanent dipole moments
have  been performed by Park {\it et al} \cite{park2000,jeung2004} and by Kotochigova {\it et al} 
\cite{kotochigova2003} using quite different theoretical approaches.
Their predictions are compared with our calculations A and B in figure \ref{fig:krb_dip}. It is striking to see that
 quite large differences are observed, up to 20\% with the non-relativistic result of ref.\cite{kotochigova2003}, and
40\% with ref.\cite{park2000} or with the relativistic calculations of ref.\cite{kotochigova2003}. It is not easy 
to understand such large differences. One reason could be the different ways to treat the core polarization effects. 
Kotochigova {\it et al} \cite{kotochigova2003} indicate that the implementation of core polarization effects may 
reduce their dipole moment by about 50\%. They also mention that their convergence with respect to the number of
basis functions is of the same order than the difference between their relativistic and non-relativistic calculations.
It would be interesting to perform these kind of comparisons for molecules with a larger dipole moment, as we may expect
larger core polarization effects. Given the good agreement we obtain with available experimental values on other systems,
we believe that our values are quite accurate.

\section{Discussion}
\label{sec:discussion}
We performed potential curves and transition dipole moments calculations for all heteronuclear alkali pairs, 
in order to investigate the influence of the basis set size on higher excited electronic states, which is expected to 
be significant. The comparison of these potential curves with other existing calculations will be presented in further
 publications, as well as results for dipole transition moments which are not available elsewhere for most species. 
These data are of particular importance for the optimization of multiple-step schemes for ultracold molecule formation, 
like the one performed by Sage {\it et al} \cite{sage2005}, or for the interpretation of recent observation of
Autler-Townes effect in highly excited molecular states \cite{garcia-fernandez2005}.
An important issue is also the evaluation of the accuracy of potential curves (by looking
for instance at well depths and equilibrium distances)
compared to experimental determinations, in relation with the choice of the effective terms included in the core-core 
potential: Table \ref{tab:other} displays the contributions of the $V_{cc}^{disp}$ term, as evaluated with equation 
\ref{eq:disp}, and of the $V_{cc}^{rep}$ term from refs.\cite{pavolini1989,jeung1997} close to the equilibrium distance of the 
ground state of the KRb and RbCs systems. The largest effect is obtained for the heaviest pair RbCs. 
Furthermore, the $V_{cc}^{disp}$ term should be cut off at a somewhat arbitrary distance, when ionic cores become close 
to each other. Jeung \cite{jeung1997} discussed various approximations for the repulsive term, leading to very different estimates, 
depending on the chosen model (see the table). Therefore it is tedious to predict well depths with an accuracy better 
than a few tens of wave numbers.
An empirical  solution to this problem is offered when an experimental determination of the ground state potential is 
available, as illustrated for example in the spectroscopic analysis of NaRb states by Docenko {\it et al}
 \cite{docenko2004}: the difference between the experimentally determined ground state potential curve and the 
computed curves of ref.\cite{korek2000} is attributed to the cumulative effect of the different terms of the 
core-core interaction. This difference is then assumed to be independent of the molecular electronic excitation: it can 
be added in turn to the computed excited potential curves, yielding then a good representation of the excited
molecular states.  

\begin{acknowledgments}
Enlightening discussions with Fernand Spiegelmann from LCAR in Toulouse (France) are gratefully acknowledged. 
We would like also to thank Philippe Milli\'e for his support at the early stage of this work.
This work has been performed in the framework of the european Research and Training
 Network 'Cold Molecules' (COMOL, HPRN-CT-2002-00290).
\end{acknowledgments}


\newpage

\begin{table}[h]
\center
\begin{tabular} {|c|c|c|c|} \hline
Atom & basis (A) &basis(B)&basis(C) \\ \hline
Li&   $8s7p4d1f/[7s5p3d1f]$& $10s7p5d3f$&\\
Na&   $7s5p5d2f/[5s5p3d2f]$& $8s6p5d2f$& \\
K&    $7s5p7d2f/[6s4p4d2f]$& $7s5p7d2f$& \\
Rb& $7s4p5d1f/[6s4p4d1f]$& $9s6p6d4f$& \\
Cs& $7s4p5d1f/[6s4p4d1f]$& $7s4p5d4f$&$9s6p6d4f$ \\ \hline
\end{tabular}
\caption{Gaussian basis sets for each studied alkali atom. The contractions are specified in brackets
for the "A" basis. The "C" basis is set up only for Cs.}
\label{tab:base}
\end{table}

\newpage

\begin{longtable}{|c|c|c|l|}
\caption{Exponents of the Gaussian functions introduced in the various basis A, B, C. When appropriate,
contracted orbitals are displayed in brackets, with contraction coefficients in parenthesis.}\\
\hline
Atom&Basis&$\ell$&Exponents \\ \hline
Li&A&s&[2.464(-0.013829),1.991(-0.032077)],0.582,0.2,0.07,0.031,0.015,0.007 \\
  & &p&[0.630(0.052433),0.240(0.078566)],0.098,0.043,0.02,[0.01(0.01445),0.005(0.00657)] \\
  & &d&0.2,0.07,[0.022796(0.491521),0.008574(0.549272)] \\
  & &f&0.01 \\
  &B&s&2.464,1.991,1.,0.582,0.2,0.136,0.07,0.011,0.008,0.004 \\
  & &p&0.630,0.24,0.098,0.03,0.02,0.01,0.004 \\
  & &d&0.2,0.062,0.024,0.01,0.004 \\
  & &f&0.01,0.0125,0.04 \\\hline
Na&A&s&[2.8357(0.007043),0.49318(-0.1871),0.072085(0.3679)],0.036061,0.016674,0.00693,0.00287 \\
  & &p&[0.431(-0.01778),0.09276(0.2003)],0.03562,0.01447,0.0058,0.00023 \\
  & &d&[0.292(0.01454),0.06361(0.123)],0.02273,0.008852,0.00352 \\
  & &f& 0.015,0.0055 \\
  &B&s&2.8357,1.,0.49318,0.072085,0.036061,0.017674,0.00693,0.00287 \\
  & &p&0.431,0.09276,0.03562,0.1447,0.0058,0.00023 \\
  & &d&0.292,0.06361,0.02273,0.008852,0.00352 \\
  & &f&0.015,0.0055 \\ \hline
 K&A&s&[0.9312(0.02463),0.2676(-0.2628)],0.0417,0.02815,0.01448,0.0055,0.0026 \\
  & &p&0.133,0.05128,0.01642,0.0052,0.0022 \\
  & &d&[1.255(0.02754),0.4432(0.05391),0.109(0.1083)],0.02994,0.01013,0.0037,0.0018 \\
  & &f&0.015,.005 \\
  &B&s&0.9312,0.2676,0.051,0.02715,0.01448,0.0055,0.0026 \\
  & &p&0.134,0.05328,0.01642,0.0053,0.002 \\
  & &d&1.255,0.443,0.108,0.02994,0.01013,0.0037,0.0018 \\
  & &f&0.015,0.005  \\ \hline

%
Rb&A&s&[1.292561(-0.065450),0.824992(0.037862)],0.234682,0.032072,0.013962,0.005750,0.0025 \\
  & &p&0.128,0.040097,0.014261,0.004850 \\
  & &d&[0.408807(0.0943),0.096036(0.1846)],0.026807,0.009551,0.0004 \\
  & &f&0.2 \\
  &B&s&2.5,1.292561,0.824992,0.5,0.234682,0.032072,0.013962,0.00575,0.025 \\
  & &p&0.25,0.128,0.040097,0.03,0.014261,0.00485 \\
  & &d&0.328807,0.2,0.096036,0.026807,0.009551,0.0004 \\
  & &f&0.2,0.1,0.05,0.005  \\ \hline
Cs&A&s&[0.328926(0.411589),0.241529(-0.682422)],0.050502,0.029302,0.013282,0.00528,0.003 \\
  & &p&0.12,0.0655,0.0162,0.00443 \\
  & &d&[0.196894(0.18965),0.067471(0.22724)],0.027948,0.010712,0.003 \\
  & &f&0.1 \\
  &B&s&0.328926,0.221529,0.050502,0.029302,0.013282,0.00528,0.003 \\
  & &p&0.155,0.0655,0.0162,0.00443 \\
  & &d&0.18894,0.067471,0.027948,0.010712,0.003 \\
  & &f&0.05,0.025,0.0125,0.005  \\ 
  &C&s&2.5,0.328941,0.241529,0.1,0.050502,0.029302,0.012,0.005,0.002 \\
  & &p&0.25,0.12,0.0655,0.03,0.012,0.005 \\
  & &d&0.3,0.196894,0.09,0.03 \\
  & &f&0.05,0.025,0.0125,0.005  \\ \hline
\label{tab:exponents-cont}
\end{longtable}

\newpage

\begin{table}[h]
\center
\begin{tabular} {|c|c|c|c|c|c|c|} \hline
Atom&Basis& $\alpha_d^{M^+}$&$\rho_s$&$\rho_p$&$\rho_d$&$\rho_f$\\ \hline
Li&A&0.1915\cite{muller1984}&1.434&0.979&0.6&0.4\\
  &B&0.1997\cite{wilson1970}&1.315&0.999&0.6&0.9\\ \hline
Na&A&0.9947\cite{muller1984}&1.42&1.625&1.5&1.5\\
  &B&0.9987\cite{wilson1970}&1.456&1.467&1.5&1.5\\ \hline
 K&A&5.354\cite{muller1984}& 2.067&1.905&1.96&1.96\\
  &B&5.472\cite{wilson1970}& 2.11439&1.97355&1.9884&1.9884\\ \hline
Rb&A&9.245&2.513&2.279&2.511&2.511\\
  &B&9.245\cite{wilson1970}&2.5538&2.349&2.5098&2.5098\\ \hline
Cs&A&15.117\cite{pavolini1989}&2.6915&1.8505&2.807&2.807\\
  &B&16.33\cite{wilson1970,coker1976}&2.8478&1.981&2.904&2.904 \\
  &C&16.33\cite{wilson1970,coker1976}&2.8864&2.7129&2.8963&2.8963\\ \hline
\end{tabular}
\caption{Dipole polarizabilities $\alpha_d^{M^+}$ and semiempirical cutoff parameters $\rho_{\ell}$ 
introduced in the basis A, B, C. For previously published basis (A), the authors used the theoretical
values quoted in the quoted references for $\alpha_d^{M^+}$, while for basis (B) we used instead the experimental values
of Wilson {\it et al} \cite{wilson1970}. Note that for Cs, Allouche and coworkers \cite{allouche2000,korek2000} 
referred to Wilson {\it et al}, but do not actually use his value.}
\label{tab:cutoff}
\end {table}

\newpage

\begin{table}[h]
\center
\begin{tabular} {|c|c|c|c|c|c|c|c|c|c|c|c|} \hline
&&$\Delta_A$&$\Delta_B$&&&$\Delta_A$&$\Delta_B$&&&$\Delta_A$&$\Delta_B$ \\ \hline
Li&2s&5.67&0.55&Na&3s&-2.26&0.00&K&4s&0.10&-0.18 \\
&2p&0.21&-0.76&&3p&-1.46&0.47&&4p&0.03&0.14 \\
&3s&7.79&26.67&&4s&19.81&2.50&&5s&-5.49&-9.99 \\
&3p&1.80&8.47&&3d&0.64&0.01&&3d&0.58&-0.61 \\
&3d&7.94&0.20&&4p&3.94&2.67&&5p&47.01&39.38 \\
&4s&1684.04&54.70&&5s&11.75&7.54&&4d&23.07&21.30 \\
&4p&4069.42&236.05&&4d&13.70&13.39&&6s&3.01&0.01 \\
&4d&&40.36&&4f&&57.45&&4f&46.32&46.30 \\
&4f&721.84&51.28&&6p&50.12&59.90&&&& \\ \hline \hline
&&$\Delta_A$&$\Delta_B$&&&$\Delta_A$&$\Delta_B$&$\Delta_C$ \\ \hline
Rb&5s&0.08&-0.03&Cs&6s&0.32&0.15&0.31&&& \\
&5p&0.00&0.00&&6p&0.05&-0.85&0.10&&& \\
&4d&0.13&0.03&&5d&-18.85&-0.51&0.46&&& \\
&6s&-32.17&-32.56&&7s&1.50&10.40&7.09&&& \\
&6p&32.67&22.92&&7p&68.43&91.63&17.40&&& \\
&5d&7.29&4.15&&6d&186.58&203.32&85.30&&& \\
&7s&-10.39&-10.79&&8s&37.55&41.02&20.20&&& \\
&&&&&4f&&23.64&23.47&&& \\\hline \hline
\end{tabular}
\caption{Differences (in cm$^{-1}$) between computed and experimental atomic binding energies for A, B, and C 
basis sets. Experimental values are taken from refs.\cite{moore1949,moore1952,moore1958}.}
\label{tab:atomic}
\end {table}

\newpage

\begin{table} [h]
\begin{tabular} {|c|c|c|c|c|c|c|c|c|c|c|}         
\hline
Molecule & LiNa & LiK & LiRb & LiCs & NaK & NaRb & NaCs& KRb & KCs & RbCs \\ \hline
basis (A)& 98   & 104 & 89   & 89   & 114 & 99   & 99  & 105 & 105 & 90   \\
basis (B)& 142  & 148 & 162  & 149  & 136 & 143  & 137 & 156 & 143 & 157  \\
basis (C)&  -   &  -  &  -   & 162  &  -  &  -   & 156 &  -  & 156 & 170  \\ \hline
\end{tabular}
\vspace{.5cm}
\caption{ Size of the  molecular basis set generated from method A, B and C.}
\label{tab:base2}
\end{table}

\newpage

\begin{longtable} {|c|ccc|cc|c|ccc|ccc|}
\caption{Computed permanent dipole moments $\vert D \vert$ of the $X^1\Sigma^+$ state (in Debye) at
equilibrium distances $R_e^X$, at the minimum distance $R_d^X$ (in a.u.), and for the $v=0$ level of the $X$ state.
Our results are compared to available experimental and theoretical values.}\\
\hline 
&\multicolumn{6}{|c|}{This work}&\multicolumn{3}{|c|}{Exp.}&\multicolumn{3}{|c|}{Th}\\ \hline
&&$D(R_e^X)$&$R_e^X$&$D(R_d^X)$&$R_d^X$&$D(v=0)$&$D$&$R_e$&ref.&$D(R_e)$&$R_e$&ref. \\ \hline
LiNa&(A)&0.561&5.43&0.630&6.85&0.566&0.45( )&    &\cite{engelke1982}&1.24&5.64&\cite{bertoncini1970}\\
    &(B)&0.554&5.42&0.633&6.89&0.556&0.47(3)&5.33&\cite{dagdigian1971}&0.485&5.429&\cite{rosmus1976}\\
    &   &     &    &     &    &     &0.463(10)&5.31&\cite{dagdigian1972}&0.485&5.42&\cite{habitz1977}\\
    &   &     &    &     &    &     &0.45     &     &\cite{tarnovsky1993}&0.487&5.47&\cite{muller1984} \\
    &   &     &    &     &    &     &         &     &                    &0.53&5.42&\cite{igel-mann1986} \\ \hline
LiK &(A)&3.558&6.21&3.807&7.50&3.565&3.510(5)&     &\cite{engelke1984}&3.437&6.292&\cite{muller1984}\\
    &(B)&3.533&6.21&3.792&7.49&3.555&3.45(2)&6.18 &\cite{dagdigian1972}&3.50&6.25&\cite{igel-mann1986}\\
    &     &    &   &     &    &     &3.87           &     &\cite{tarnovsky1993}&&& \\ \hline
LiRb&(A)&4.168&6.52&4.442&7.78&4.165&4.05           &     &\cite{tarnovsky1993}&4.13&6.52&\cite{igel-mann1986}\\
    &(B)&4.142&6.48&4.414&7.78&4.131&&&&&&\\ \hline
LiCs&(A)&5.520&6.81&6.023&8.33&5.529&6.30           &     &\cite{tarnovsky1993}&5.48&6.89&\cite{igel-mann1986}\\
    &(B)&5.512&6.82&5.998&8.31&5.524&&&&&&\\
    &(C)&5.462&6.81&5.970&8.33&5.478&&&&&&\\ \hline
NaK &(A)&2.760&6.50&2.854&7.42&2.759&2.73(9)        &     &\cite{wormsbecher1981}&3.6&6.9 &\cite{janoschek1978} \\
    &(B)&2.763&6.49&2.862&7.47&2.762&2.690(14)&6.55 &\cite{dagdigian1972}&2.735&6.36&\cite{muller1984}\\
    &     &    &   &     &    &     &3.42           &     &\cite{tarnovsky1993}&2.95 &6.614&\cite{stevens1984}\\
    &     &    &   &     &    &     &           &   &     &2.54&6.59&\cite{igel-mann1986} \\  
    &     &    &   &     &    &     &           &   &     &2.78&6.6 &\cite{magnier2000}\\ \hline
NaRb&(A)&3.304&6.84&3.413&7.75&3.306&3.10(3)   &6.73 &\cite{dagdigian1972}&3.33&6.85&\cite{igel-mann1986} \\
    &(B)&3.301&6.84&3.413&7.76&3.301&3.51           &     &\cite{tarnovsky1993}&&& \\ \hline
NaCs&(A)&4.613&7.20&4.821&8.26&4.607&4.75(20)& 6.91&\cite{dagdigian1972}&4.60&7.23&\cite{igel-mann1986} \\
    &(B)&4.661&7.20&4.864&8.26&4.660&4.75           &     &\cite{landolt1974}&&& \\
    &(C)&4.580&7.20&4.793&8.29&4.579&5.86           &     &\cite{tarnovsky1993}&&& \\ \hline


KRb &(A)&0.615&7.64&0.620&8.20&0.615&0.20           &     &\cite{tarnovsky1993}&0.64&7.65&\cite{igel-mann1986} \\
    &(B)&0.589&7.64&0.605&8.49&0.589&               &     &                    &0.71&7.7&\cite{kotochigova2003}\\
    &     &    &   &     &    &     &               &     &                    &1.06&7.7&\cite{jeung2004} \\ \hline
KCs &(A)&1.906&8.01&1.957&8.87&1.906&2.58           &     &\cite{tarnovsky1993}&1.92&8.05&\cite{igel-mann1986} \\
    &(B)&1.921&8.02&1.967&8.85&1.921&               &     &                    &&& \\
    &(C)&1.835&8.02&1.891&8.93&1.837&               &     &                    &&& \\ \hline
RbCs&(A)&1.238&8.28&1.278&9.19&1.237&2.39           &     &\cite{tarnovsky1993}&1.26&8.71&\cite{igel-mann1986} \\
    &(B)&1.278&8.30&1.309&9.12&1.280&               &     &                    &&& \\
    &(C)&1.205&8.30&1.240&9.19&1.204&               &     &                    &&& \\ \hline \\ \\
\label{tab:compar_cont} 
\end{longtable}

\newpage

\begin{table} [h]
\begin{tabular} {|c|c|c|c|cc|} \hline
    &$R_e^X (a_0)$&$V_{cc}^{disp}$&\multicolumn{3}{|c|}{$V_{cc}^{rep}$} \\  
    &             &               &\cite{pavolini1989}&\cite{jeung1997}-A&\cite{jeung1997}-E \\ \hline
KRb &7.64         &-6.41           &-                 &$\approx$17       &$\approx$-507 \\ \hline
RbCs&8.30         &-10.22          &18.42             &$\approx$25       &$\approx$-500 \\ \hline
\end{tabular}    
\caption{Contribution (in cm$^{-1}$) of the $V_{cc}^{disp}$ and $V_{cc}^{rep}$ terms around the equilibrium distance 
$R_e^X$ of the KRb and RbCs ground state. The $V_{cc}^{disp}$ term is estimated following equation \ref{eq:disp} and 
parameters from method A. Both
two results (labelled as A and E) from ref. \cite{jeung1997} are displayed.}\label{tab:other} 
\end{table}

\newpage
\section{Figure captions}

\begin{figure}
\center\includegraphics[width=0.9\columnwidth]{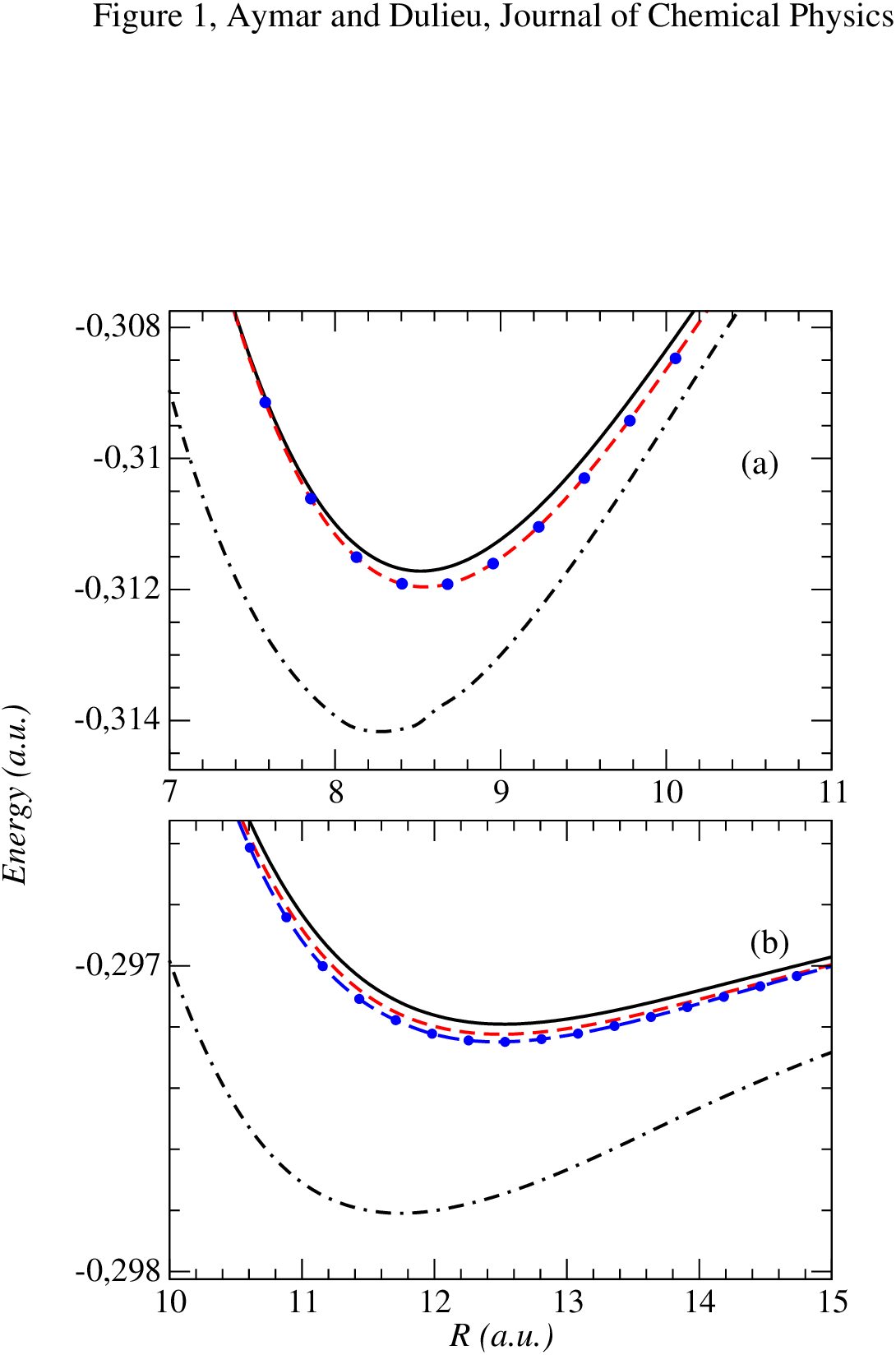}
\caption{}
\label{fig:rbcspot_Xa}
\end {figure}

Figure \ref{fig:rbcspot_Xa}: 
({\it Color on line.}) The potential curves $V'(R)$ (including only the $1/R$ term of $V_{cc}(R)$) around their minimum
for 
(a) the $X^1\Sigma^+$ and (b) the $a^3\Sigma^+$ states of RbCs, as obtained through the three approaches A 
(black full lines), B (red dashed lines), C (blue dashed line with dots), described in the text.
 In (a), results for B and C approaches are superimposed. In addition, the black dot-dashed lines represent 
$V'(R)+V_{cc}^{ind}(R)$ for both states, in the A case.

\begin{figure}
\center\includegraphics[width=0.9\columnwidth]{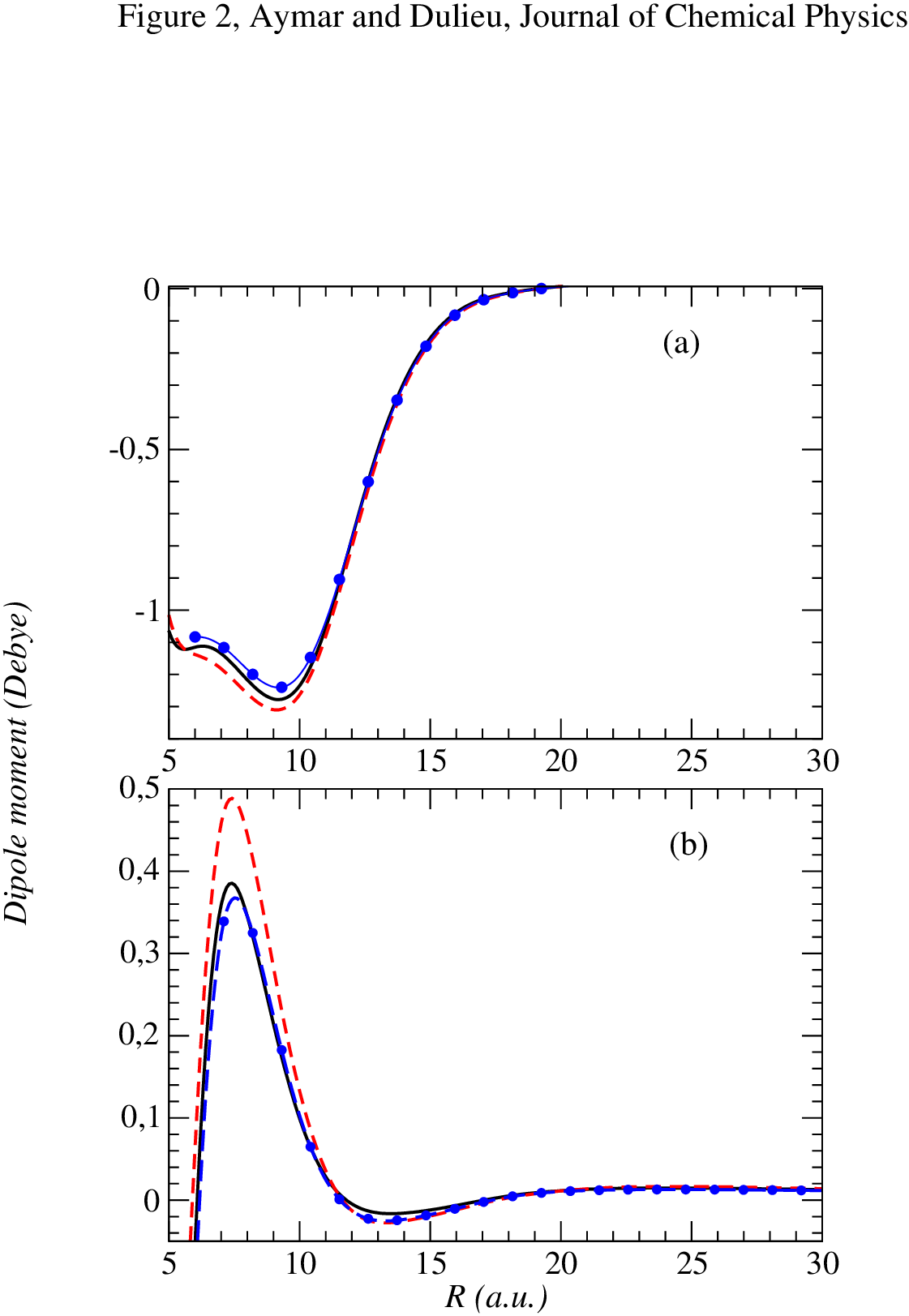}
\caption{}
\label{fig:rbcsdip_Xa}
\end {figure}

Figure \ref{fig:rbcsdip_Xa}:
({\it Color on line.}) The permanent electric dipole moments (in Debye) of (a) the $X^1\Sigma^+$ (b) the $a^3\Sigma^+$ states,
in RbCs, as obtained through the three approaches A, B, C, described in the text.

\begin {figure}
\center\includegraphics[width=0.9\columnwidth]{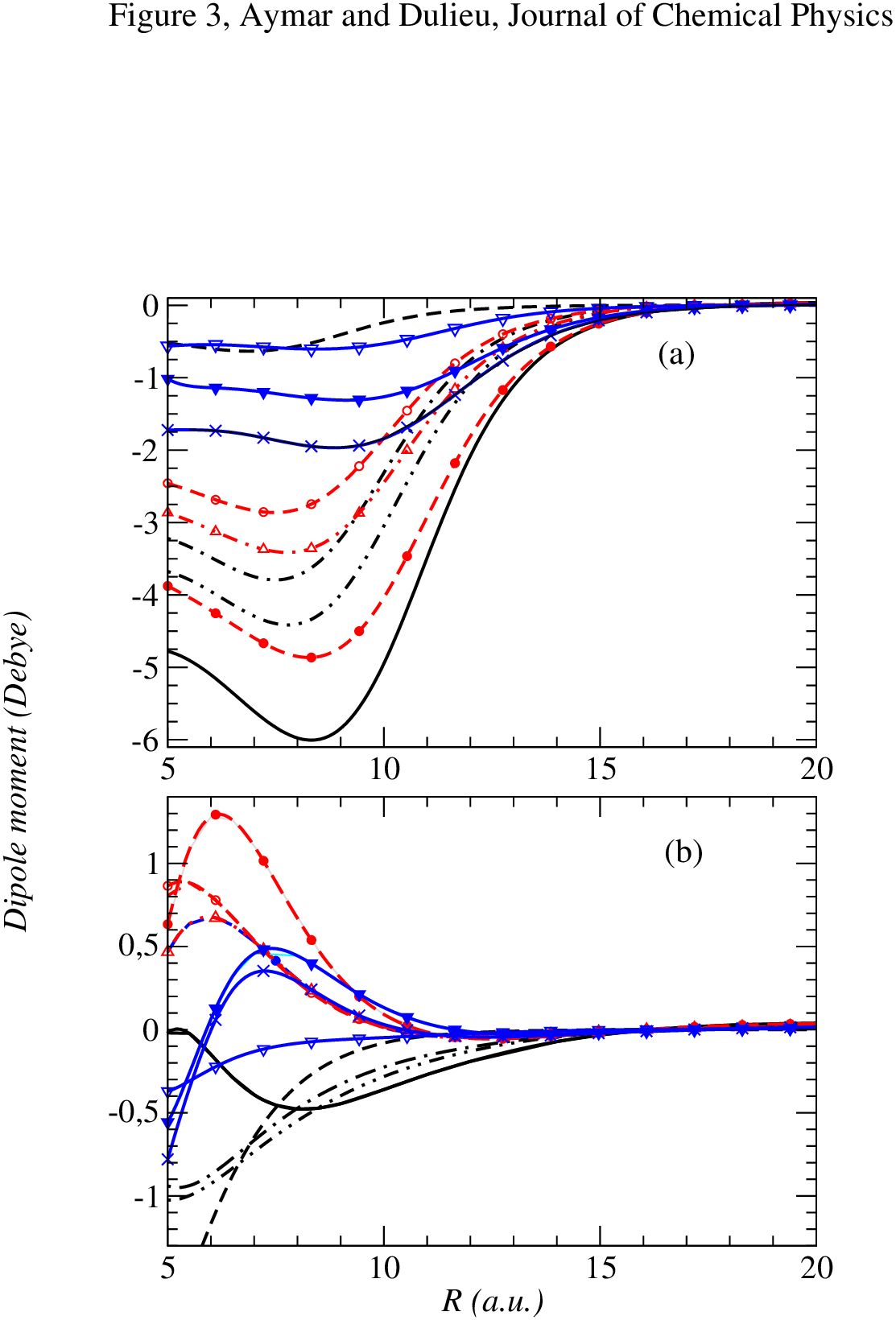}
\caption{}
\label{fig:dipole_R}
\end {figure}

Figure \ref{fig:dipole_R}:
({\it Color on line.}) The permanent electric dipole moments (in Debye) of (a) the $X^1\Sigma^+$ state (b) the $a^3\Sigma^+$ state,
as functions of the internuclear distance, obtained from calculation B. Results are displayed in black for Li coumpounds 
(dashed line for LiNa, dot-dashed line for LiK, double-dot-dashed line for LiRb, full line fo LiCs), in red for Na
compounds (open circles for NaK, upper triangles for NaRb, closed circles for NaCs), and in blue for K compounds (lower
open triangles for KRb, crosses for KCs) and for RbCs (lower closed triangles).

\begin {figure}
\center\includegraphics[width=0.9\columnwidth]{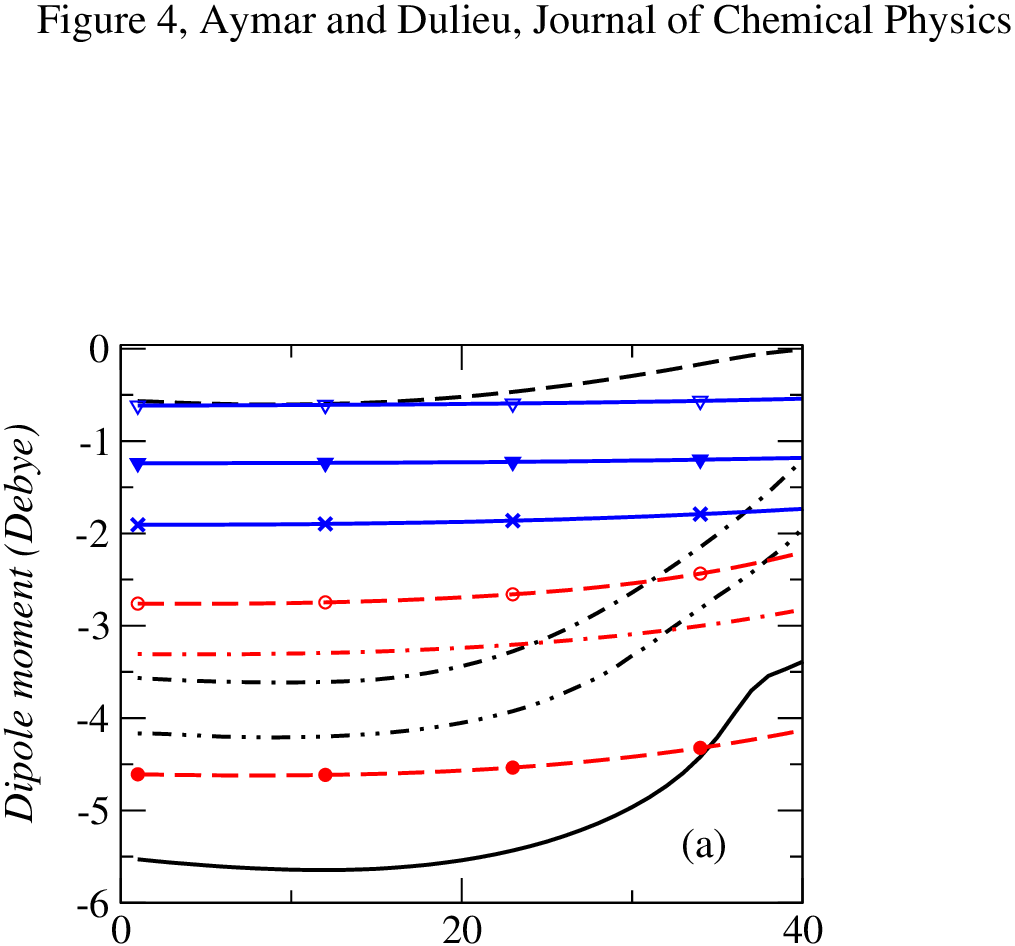}
\caption{}
\label{fig:dipole_v}
\end {figure}

Figure \ref{fig:dipole_v}:
({\it Color on line.}) The permanent electric dipole moments (in Debye) of the $X^1\Sigma^+$ state
of the mixed alkali pairs as functions of the vibrational level, obtained from calculations B. The color and line code
is the same than in figure \ref{fig:dipole_R}.

\begin {figure}
\center\includegraphics[width=0.9\columnwidth]{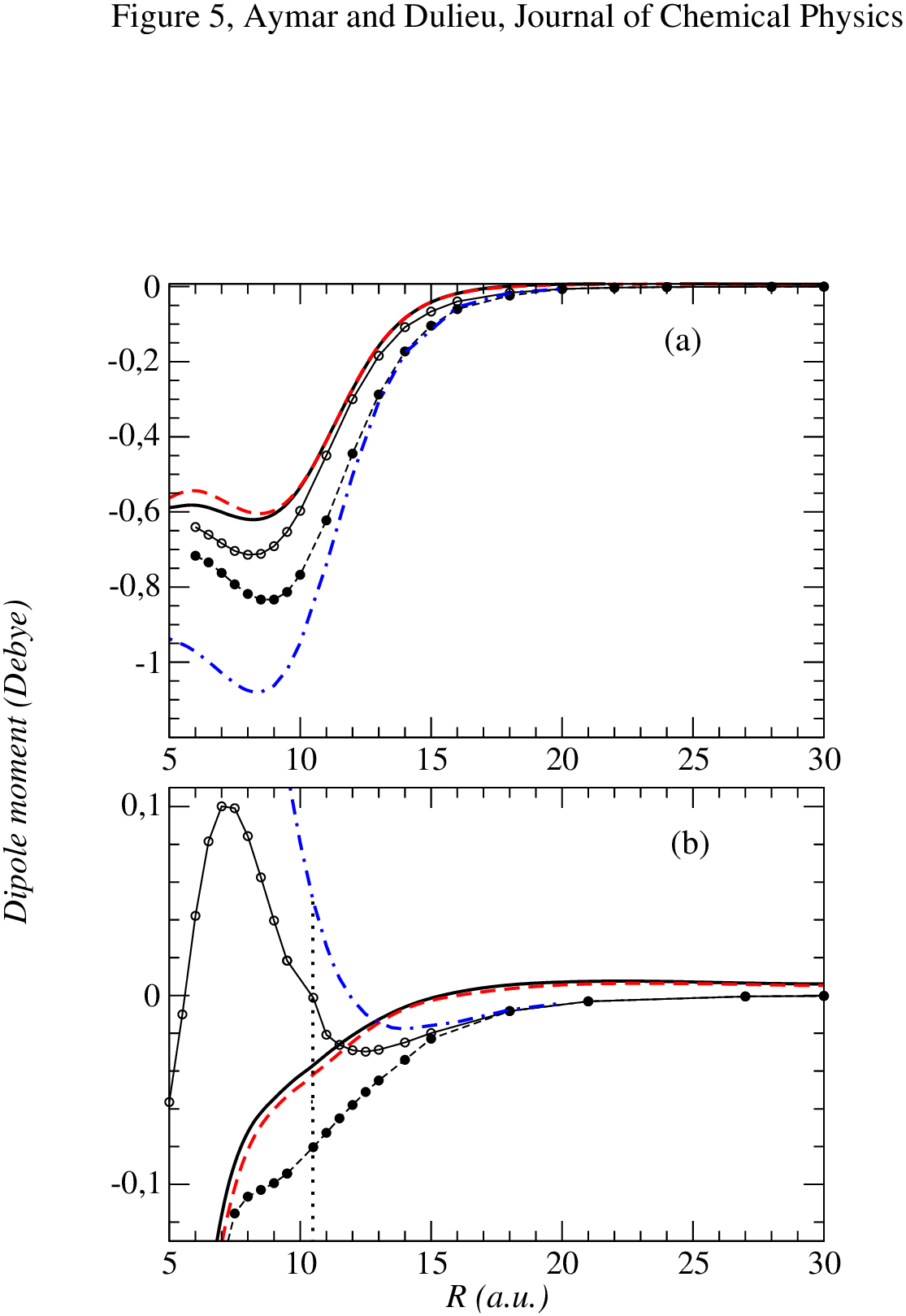}
\caption{}
\label{fig:krb_dip}
\end {figure}

Figure \ref{fig:krb_dip}:
({\it Color on line.}) The permanent electric dipole moments (in Debye) of (a) the $X^1\Sigma^+$ state (b) the $a^3\Sigma^+$ state 
of KRb as functions of the internuclear distance, obtained from calculations A (full line) and B (dashed line). Our
results are compared to other available theoretical determinations: non-relativistic (open circles) and
relativistic (closed circles) results from ref.\cite{kotochigova2003}, and non-relativistic results from 
ref.\cite{park2000,jeung2004} (dot-dahed blue line). The vertical dashed line in (b) panel indicates the location 
of the repulsive wall of the potential at the energy of its dissociation limit.

\newpage

\newpage

\newpage

\newpage

\newpage

\end{document}